# Transferable potential for molecular dynamics simulations of borosilicate glasses and structural comparison of machine learning optimized parameters

Kai Yang[a, b], Ruoxia Chen[a], Anders K.R. Christensen[c], Mathieu Bauchy[a], N.M. Anoop Krishnan[d], Morten M. Smedskjaer[c], Fabian Rosner[a*]

[a] Department of Civil and Environmental Engineering, University of California, Los Angeles, CA 90095, USA

[b] Department of Materials Science and Engineering, University of California, Los Angeles, CA 90095, USA

[c] Department of Chemistry and Bioscience, Aalborg University, 9220 Aalborg, Denmark

[d] Department of Civil Engineering, Indian Institute of Technology Delhi, Hauz Khas, New Delhi 110016, India

\* Corresponding author.

Email address: fabianrosner@g.ucla.edu. 5731-H Boelter Hall, Los Angeles, CA 90095-1593. Phone number: +1 (310)825-2471

## Abstract

The simulation of borosilicate glasses is challenging due to the composition and temperature dependent coordination state of boron atoms. Here, we present a newly developed machine learning optimized classical potential for molecular dynamics simulations that achieves transferability across diverse borosilicate glass compositions. Our potential accurately predicts the glass structural variations in short- and medium-range order in different glass compositions, including validating our potential against experimental X-ray structure factor data. Notably, these data are not included in the optimization framework, which focuses exclusively on density and four-fold coordinated boron fraction. We further investigate the impact of empirical parameters in the force field formulation on the microscopic bond lengths, bond angles and the macroscopic densities, providing new insights into the relationship between interatomic potentials and bulk glass behaviors.

## 1. Introduction

Borosilicate glasses are widely used across diverse applications, including optical fibers, laboratory equipments, pharmaceutical packaging, and nuclear waste immobilization[1–4], owing to their exceptional properties such as low thermal expansion, high thermal shock resistance, and superior chemical durability[2,5]. Detailed structural information of these glasses can offer a better understanding of the composition-structure-property relationship, in turn assisting the design of new glass compositions. However, the atomic structure of borosilicate glasses is inherently complex and composition-dependent, primarily due to the variable coordination environment of boron atoms, which evolves with composition, thermal and pressure history[6]. The coexistence of trigonal $BO_3$ and tetrahedral $BO_4$ units within the borosilicate network creates structural complexity that remains challenging to characterize[7].

While molecular dynamics (MD) simulations have successfully enhanced our understanding of numerous silicate systems at the atomic scale[8,9], borosilicate glasses remain a formidable computational challenge. Recently proposed interatomic potentials, such as Kieu's potential[10], have improved certain aspects but suffer from significant limitations, including the need to recalibrate multiple parameters for different compositions, limited predictive capability, and poor transferability across glass systems. Wang *et al.* developed a transferable potential for borosilicate glasses[11], but a subsequent study has revealed reproducibility issues attributed to incorrect atomic masses, resulting in inaccurate glass density predictions[12]. Deng and Du also proposed a set of empirical potentials for multicomponent borosilicate glasses[13], adopted from Teter's parameterization[14]. However, traditional approaches for developing classical MD potentials still rely on trial-and-error methods or empirical fitting procedures, which are often inefficient for high dimensional parametrization and introduce systematic biases that compromise model reliability.

Machine learning has emerged as a transformative solution, offering sophisticated optimization algorithms that can systematically navigate high-dimensional parameter landscapes with enhanced efficiency and objectivity. Recent investigations have shown that ML-based methodologies can efficiently develop interatomic forcefields achieving exceptional agreement between computational predictions and experimental observations[15–18]. Building on this foundation, we have recently developed an active learning framework for parameterizing borosilicate glass potentials using experimental density and boron coordination data as primary optimization objectives. Our active learning methodology navigates the parameter space by systematically selecting the most informative training configurations, thereby enhancing learning effectiveness while reducing computational demands. More detailed information on the algorithm and optimization procedure can be found in Ref. [19].

Here, to improve the inaccuracies of Wang's potential, we report a ML-parametrized transferable potential for borosilicate glasses. While being a classical potential, it is derived from our ML optimization framework, which uses experimental densities and four-fold coordinated boron fractions data as optimization targets. The resulting potential demonstrates excellent transferability across a wide range of borosilicate compositions while maintaining constant, composition-independent parameters. The simulated glasses accurately reproduce the compositional dependence of both short- and medium-range structural properties, including density, coordination numbers, pair distribution functions (PDFs), bond angle distributions (BADs), ring size distributions, and structure factors. Furthermore, we conduct a systematic analysis of how force field parameters control macroscopic glass properties, revealing the underlying mechanisms that link atomic-level interactions to bulk glass behavior.

## 2. Method

### Molecular dynamics simulation

All borosilicate glass systems, with their compositions listed in Table. 1, were simulated using molecular dynamics simulations implemented in the Large-scale Atomic/Molecular Massively Parallel Simulator (LAMMPS) software[20]. To ensure reproducibility and provide uncertainty estimates, we performed six independent simulations for each composition using the same process but different initial configurations. Our simulations employed an 11 Å cutoff for both short-range and Coulombic interactions, with long-range electrostatic forces computed via the particle-particle particle-mesh (PPPM) algorithm maintaining $10^{-5}$ accuracy. Integration was performed using a 1.0 fs timestep to ensure numerical stability and energy conservation.

**Table 1.** Compositions (in mol%) of the simulated glasses herein.

| Sampled ID | $SiO_2$ | $B_2O_3$ | $Na_2O$ | CaO |
|---|---|---|---|---|
| 75B | 0 | 75 | 15 | 10 |
| 62B | 13 | 62 | 15 | 10 |
| 50B | 25 | 50 | 15 | 10 |
| 37B | 38 | 37 | 15 | 10 |
| 24B | 51 | 24 | 15 | 10 |
| 12B | 63 | 12 | 15 | 10 |
| 6B | 69 | 6 | 15 | 10 |
| 0B | 75 | 0 | 15 | 10 |
| 10B [a] | 60 | 10 | 15 | 15 |

[a] This composition was not used for parametrization, but has been simulated to compare with neutron diffraction data from literature [6].

Glass structures were prepared through a controlled melt-quench protocol designed to eliminate structural memory effects. Initially, approximately 3000 atoms were randomly distributed within cubic simulation cells while preventing unphysical atomic overlaps. Each system underwent complete structural equilibration through a two-stage melting process: 10 ps at 3000 K under canonical (*NVT*) conditions, followed by 100 ps under isothermal-isobaric (*NPT*) conditions at zero pressure, ensuring complete erasure of initial configurational bias. Subsequently, all systems were subjected to controlled linear cooling from 3000 to 300 K at 1 K/ps under *NPT* conditions, qualitatively replicating experimental quenching protocols. Final glass configurations were equilibrated through 100 ps *NPT* relaxation at 300 K and zero pressure, followed by 100 ps *NVT* production runs for statistical analysis.

All reported glassy-state properties represent ensemble averages computed over 100 independent configurations extracted at 1 ps intervals during production runs, ensuring robust statistical sampling. To address occasional structural instabilities manifesting as volume divergence during high-temperature *NPT* equilibration, we implemented a pre-conditioning protocol, involving initial *NVT* melting (100 ps at 3000 K) followed by isochoric cooling to 300 K, generating stable starting configurations for subsequent melt-quench procedures.

## Machine learning optimized parametrization

We developed an active learning-based optimization framework to calibrate classical interatomic potentials for borosilicate glasses directly against experimental data. The process proceeds iteratively. At each iteration, MD simulation results are used as ground truth to train a multi-layer perceptron (MLP) surrogate model that approximates the MD behavior. The MLP predicts glass density and fraction of 4-fold boron ($B^4$) species as functions of both potential parameters and glass composition. By explicitly including the $B_2O_3$ molar fraction as an input, the model becomes composition-aware and can generalize across a wide range of borosilicate glass compositions. Based on this surrogate model, an optimization algorithm proposes candidate parameter sets, which are evaluated by the surrogate. The most promising candidates are then selected for full MD simulations, and the resulting data are incorporated into the training set to refine the surrogate model in the next iteration. This active learning loop enables efficient exploration of the parameter space by focusing on the most promising regions, significantly reducing the number of required MD simulations.

To efficiently explore the five-dimensional parameter space, we initially employed Bayesian optimization and later tested the Covariance Matrix Adaptation Evolution Strategy (CMA-ES). BO leverages a probabilistic surrogate and an acquisition function to balance exploration and exploitation, while CMA-ES samples candidates from a multivariate distribution and adaptively updates its search strategy based on performance. This makes it particularly effective for broad and noisy search spaces. In each optimization iteration, multiple runs of the algorithm are executed in parallel, and the top ten high-performing parameter sets—based on predicted deviations in density and $B^4$ fraction from experimental targets—are selected for MD validation. This adaptive and data-efficient framework accelerates convergence toward transferable and physically accurate potential parameters across diverse glass compositions. Further details about the optimization workflow are provided in our recent related paper[19].

### Ring size distribution

We employed the RINGS package[21] to compute the ring size distribution of the simulated glasses. In this analysis, rings are defined as the shortest closed paths within the borosilicate network, with their size corresponding to the number of Si or B atoms that constitute each ring. 2.0 Å cutoff is set for computing B–O and Si–O bonds in the RINGS input file.

### Experimental X-ray structure factor

For structural validation of the trained potentials, we performed X-ray structure factor $S(Q)$ measurements on glasses of the same compositions as those used in the simulations. These samples were taken from a previous study[22], but only the six compositions from 6B to 62B were available (i.e., not the two end-member glasses). Then, X-ray total scattering measurements were performed using a laboratory diffractometer (STOE STADI P) equipped with an Ag X-ray source monochromatized to only emitting K$\alpha_1$ radiation ($\lambda$ = 0.559407 Å). The samples were crushed and filled in special glass capillaries with an outer diameter of 1 mm (Capillary Tube Supplies Ltd) and then flame shut to prevent moisture uptake. Data was collected with two MYTHEN 2K detectors in the Debye-Scherrer mode in a range 2θ of 1.350 to 134.025 with a total measuring time of 24 h, resulting in a total scattering $S(Q)$ dataset with $Q_{max}$ of 19.5 Å$^{-1}$. The data was processed by GudrunX[23] and normalized using the analyzed compositions and densities reported elsewhere[22].

## 3. Results

### Interatomic potential parameters

We present our ML parametrized, transferable interatomic potential for molecular dynamics simulations of borosilicate glasses in Table 2. The corresponding partial charges to each element are shown in Table 3, which are adopted from the Wang *et al.*[11] and the Guillot-Sator interatomic potential[24]. As detailed in the Methods section, our parametrization focused only on tuning the parameters for B–O and B–B pairs based on a series of simulations of borosilicate glasses and corresponding experimental data[22], while other parameters were adopted from previous work[11,24].

**Tabel 2.** Parameters of the trained interatomic potential.

| Bond | $A_{ij}$ (eV) | $\rho_{ij}$ (Å) | $C_{ij}$ (eV·Å$^6$) |
|---|---|---|---|
| O–O[a] | 9022.79 | 0.265 | 85.0921 |
| Si–O[a] | 50,306.10 | 0.161 | 46.2978 |
| B–O | 191,757.12 | 0.1249 | 32.5600 |
| B–B | 532.85 | 0.3527 | 0.0 |
| Si–B[a] | 337.70 | 0.29 | 0.0 |
| Na–O[a] | 120,303.80 | 0.17 | 0.0 |
| Ca–O[a] | 155,667.70 | 0.178 | 42.2597 |

[a] These parameters are taken from the Wang *et al.* potential[11].

**Tabel 3.** Fixed partial charge attributed to each element.

| Element | Partial charge (e) |
|---|---|
| O | -0.945 |
| Si | 1.89 |
| B | 1.4175 |
| Na | 0.4725 |
| Ca | 0.945 |

## Density and boron speciation

We compare the densities and 4-fold coordinated boron fractions of simulated glasses using our new potential against the experimental data[22]. Figure 1demonstrates excellent agreement between simulated results using our ML-optimized potential and experimental data, accurately capturing composition trends in both density and B$^4$ fraction.

The density behavior of borosilicate glasses exhibits a non-monotonic trend with boron content (Fig. 1a), i.e., density increases with boron substitution for silicon at low B$_2$O$_3$ concentrations (<30 mol%), and then decreases at higher boron concentrations. This behavior reflects the underlying structural evolution. Initially, the shorter average B–O bond length in the BO$_4$ unit (around 1.48 Å, see Fig. 2b), compared to average Si–O bonds in the SiO$_4$ unit (around 1.63 Å, see Fig. 2a), creates denser tetrahedral networks that dominate the glass-forming structure. However, as B$_2$O$_3$ content increases, BO$_4$ units progressively convert to BO$_3$ trigonal units, as evidenced by the monotonic decrease in B$^4$ fraction (Fig. 1b). Despite the shorter average B–O bond length in BO$_3$ units (around 1.40 Å, see Fig. 2b), their three-fold coordination makes them inherently less dense than four-fold coordinated BO$_4$ units, resulting in the observed density reduction at high boron concentrations.

We observe minor discrepancies between simulated and experimental data, particularly for densities at low-boron content and B$^4$ fractions at high-boron content. These deviations highlight the inherent challenge of developing transferable potentials across broad composition ranges. Nevertheless, our potential successfully captures the key trends in both density and fraction of four-fold coordinated boron across the borosilicate glass series from pure silicate to pure borate end-members.

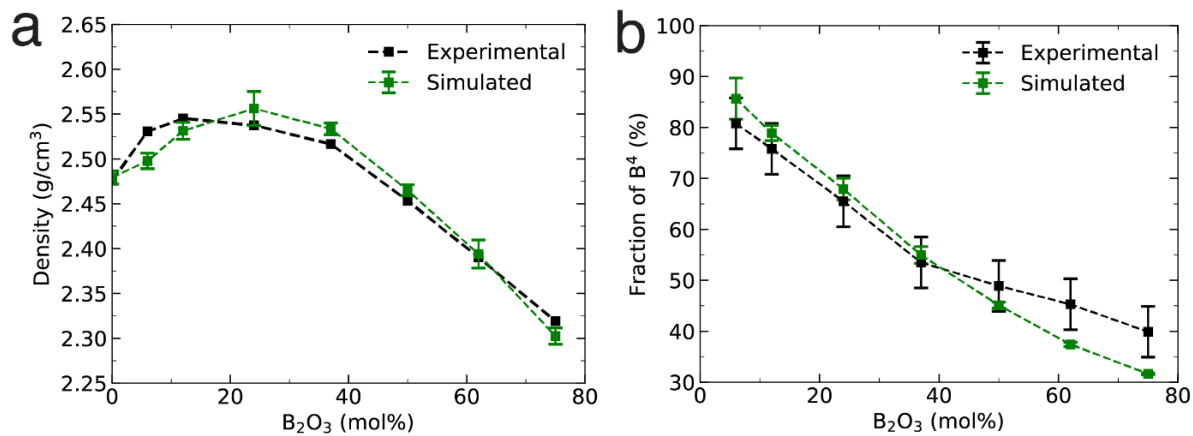

**Figure 1.** Comparison of ML-optimized potential parameters against experimental data. (a) Density and (b) fraction of four-fold coordinated boron as functions of $B_2O_3$ content. Lines serve as guides to the eye.

## Pair distribution functions

Partial pair distribution functions (PDFs) provide detailed insights into the short- and medium-range order around each element. Figure 2 presents the partial PDFs for Si–O, B–O, Ca–O, and Na–O pairs across different compositions. The local environment of Si remains largely unaffected by composition changes (Fig. 2a), as evidenced by the unchanged first coordination shell. Minor changes in the second coordination shell of Si–O pairs reflect the incorporation of boron into the network structure. In contrast, the B–O partial PDF exhibits a characteristic bimodal first peak (Fig. 2b), confirming the coexistence of $BO_3$ and $BO_4$ coordination environments. With increasing $B_2O_3$ content, the average B–O bond length decreases as $BO_3$ units become increasingly favored over $BO_4$ units, consistent with the coordination trends observed in Fig. 1b.

The network modifiers (Ca and Na atoms) show different behavior, with their local environments exhibiting high sensitivity to glass composition (Fig. 2c,d). Both Ca–O and Na–O bond distances increase substantially with $B_2O_3$ addition, reflecting a shift in cation function. That is, as $B_2O_3$ content decrease, Ca and Na cations increasingly serve as charge compensators for $BO_4$ tetrahedra rather than as traditional network modifiers. In their charge compensation role, these cations interact more strongly with oxygens, resulting in shorter bond distances compared to when they act as network modifiers.

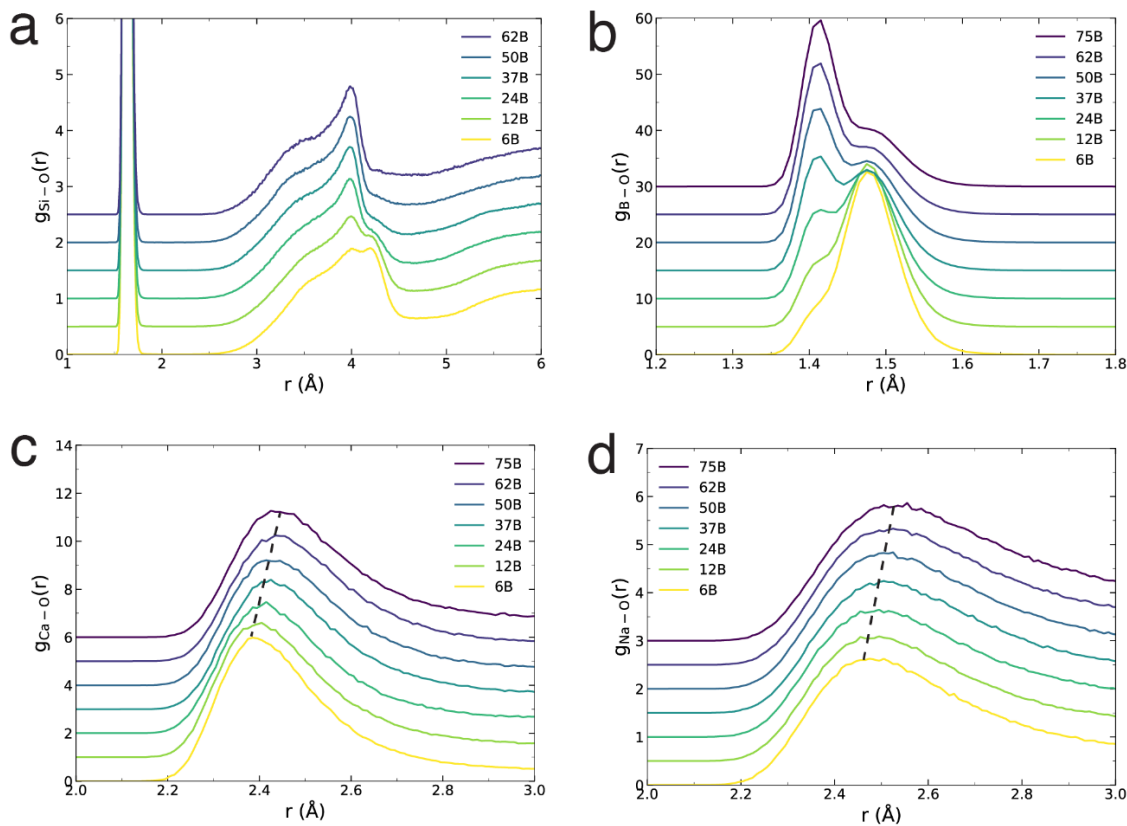

**Figure 2.** (a) Si–O, (b) B–O, (c) Ca–O, and (d) Na–O partial pair distribution functions of simulated borosilicate glasses with varying $B_2O_3$ molar fraction using the optimized potential parameters. The dashed lines are a guide for the eye.

## Bond angle distributions

To investigate medium-range order, we analyze bond angle distributions (BADs) for O–B–O, O–Si–O, and inter-polyhedral X–O–X angles (where X = Si or B) to characterize structural transitions with composition in Fig. 3. Consistent with the results in Fig. 2a, we observe that the O–Si–O tetrahedral angles remain essentially invariant with $B_2O_3$ content (Fig. 3a). Similar to the B–O partial PDFs, the O–B–O bond angle distribution exhibits a characteristic bimodal pattern, reflecting the gradual conversion from four-fold to three-fold coordinated boron with increasing $B_2O_3$ content (Fig. 3b). The average O–B–O angles center around 109° and 120° for tetrahedral $BO_4$ and trigonal $BO_3$ units, respectively, as shown in Fig. 3c, consistent with their expected geometries. Importantly, Figure 3c demonstrates that the angular environments of both $BO_3$ and $BO_4$ units also remain compositionally unaffected. In contrast, the inter-polyhedral X–O–X bridging angles show pronounced compositional dependence (Fig. 3d). These angles initially decrease with $B_2O_3$ addition, reach a minimum, and then increase at higher boron concentrations. This non-monotonic trend remarkably parallels the density behavior observed in Fig. 1a, indicating that inter-polyhedral connectivity directly influences glass density, as smaller bridging angles promote enhanced network compactness.

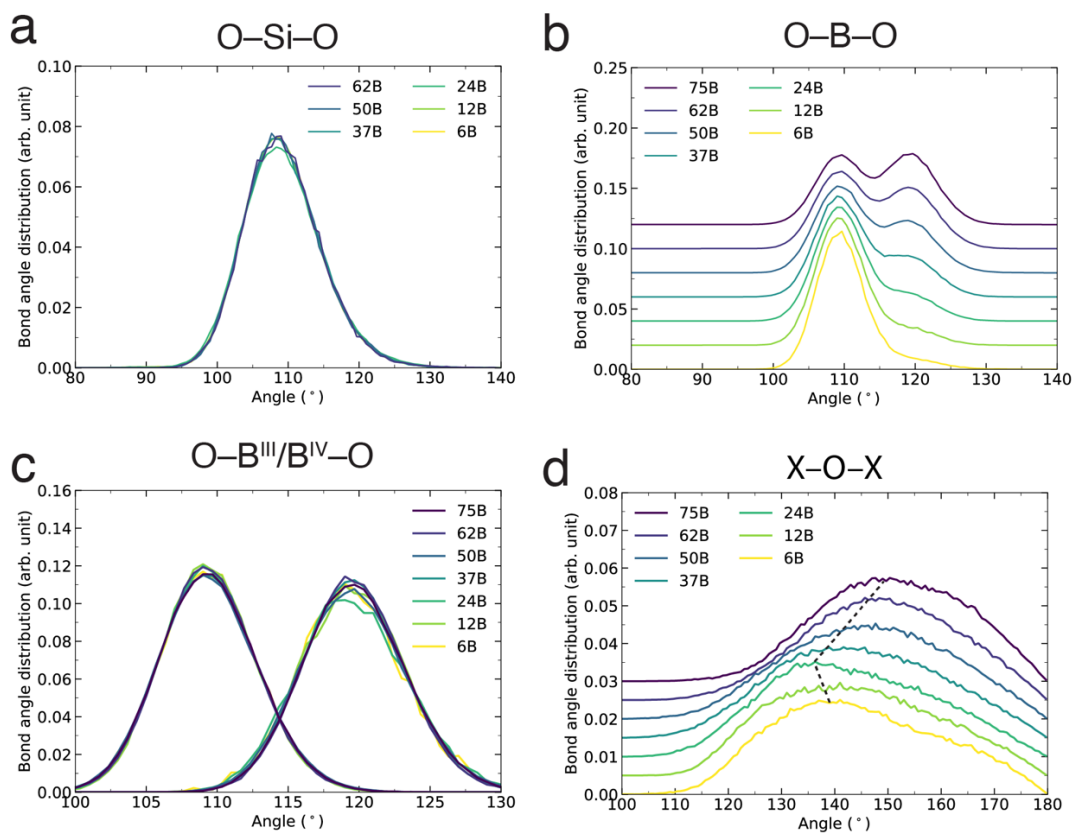

**Figure 3.** Bond angle distributions for (a) O–Si–O, (b) O–B–O, (c) O–B$^{III}$/B$^{IV}$–O, and (d) X–O–X in the simulated borosilicate glasses with varying B$_2$O$_3$ molar faction. The dashed line is a guide for the eye.

### Ring size distributions

We further investigate the medium-range order structure of the simulated glasses by analyzing ring size distributions. Figure 4 presents the computed ring size distributions, which align well with previous simulation studies[11,25,26]. The addition of B$_2$O$_3$ initially promotes the formation of smaller rings in the system, but as the B$_2$O$_3$ molar fraction continues to increase, larger rings become more prevalent (Fig. 4a). This trend is clearly illustrated in Fig. 4b, where we plot the average ring size as a function of B$_2$O$_3$ content. The average ring size exhibits a non-monotonic behavior: initially decreasing in the low-boron region, then increasing with further B$_2$O$_3$ addition. This behavior correlates with our earlier findings that the 24B composition exhibits maximum density (Fig. 1a) and minimum X–O–X inter-polyhedral bond angles (Fig. 3d). The underlying mechanism relates to network topology as smaller rings promote more compact packing.

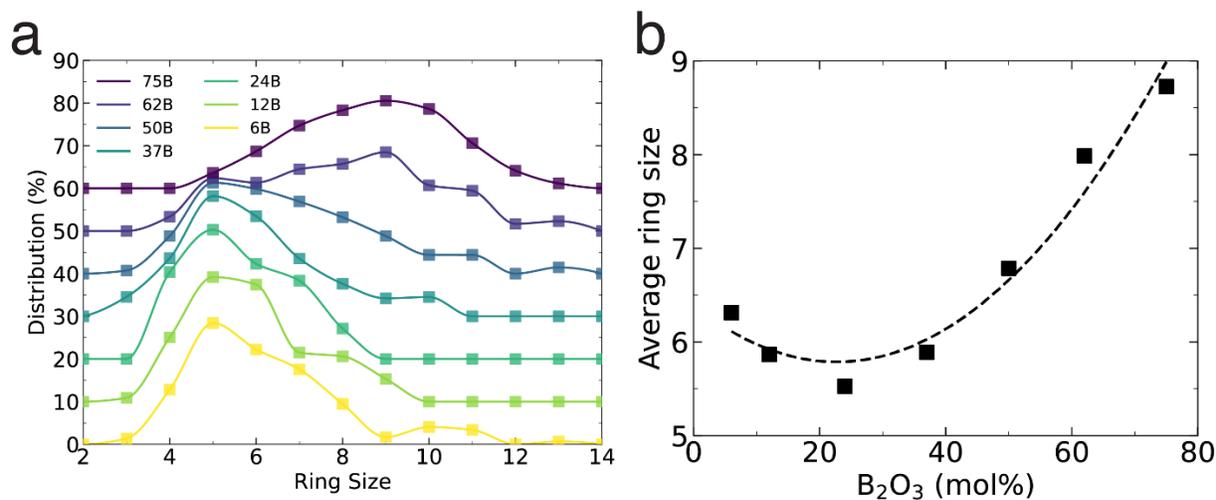

**Figure 4.** (a) Ring size distribution and (b) average ring size of simulated borosilicate glasses with different $B_2O_3$ molar fraction. The lines are guides for the eye.

### Neutron and X-ray structure factors

To validate the structural accuracy of our potential, we compare the structures of simulated glasses against experimental data. First, Figure 5 presents the total neutron structure factor from experiments in literature [6] compared against our computed values for the 10B glass. The intensities and positions of the diffraction peaks are well predicted at both low and high $Q$ values, showing that simulated glass can capture the medium- and short-range order structure of the glass, respectively. Notably, the density and $B^4$ fraction of the 10B composition were not included in our optimization targets, yet the success of reproducing the experimental neutron structure factor data confirms the transferability of our potential beyond the specific composition domain used during parameterization.

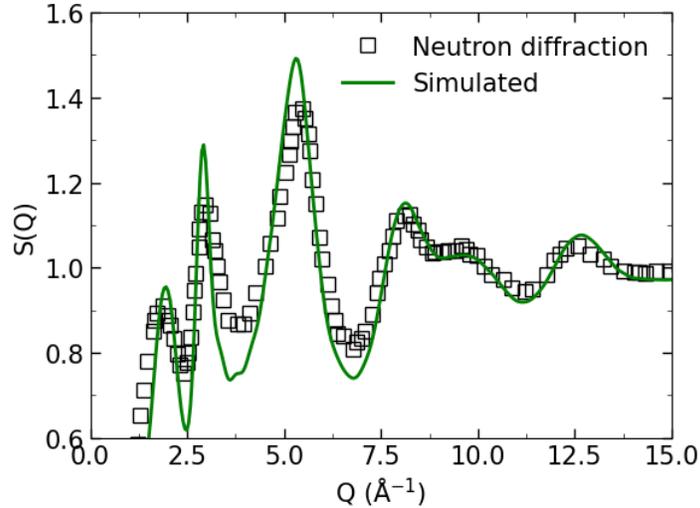

**Figure 5.** Comparison of the total neutron structure factor for 10B glass between experimental neutron diffraction data[6] and simulated glass with our potential.

To further validate our potential model, we conducted X-ray diffraction measurements on borosilicate samples spanning compositions from 6B to 62B, and compared the results with our simulated X-ray structure factors (Fig. 6). Remarkably, despite optimizing our potential using only density and boron speciation data, the simulations demonstrate good agreement with experimental diffraction patterns. That is, our model successfully reproduces the first sharp diffraction peaks across most compositions, with particularly excellent performance for low-boron content glasses. This indicates that the simulations accurately capture the medium-range structural ordering that characterizes these glass systems. For high-boron content compositions, the observed discrepancies are consistent with the $B^4$ prediction accuracy noted earlier (Fig. 1b), likely reflecting the inherent trade-offs when optimizing parameters across diverse compositional ranges[19]. The close correspondence between simulated and experimental structure factors $S(Q)$ provides strong validation of our approach. The fact that our parameter sets reproduce key structural features and overall diffraction trends, despite being constrained to match only two properties during optimization, demonstrates that the model successfully captures the fundamental physics governing borosilicate glass formation. While some discrepancies remain, particularly in the high-boron glass , the overall structural fidelity achieved is encouraging and supports the robustness of our parameterization strategy.

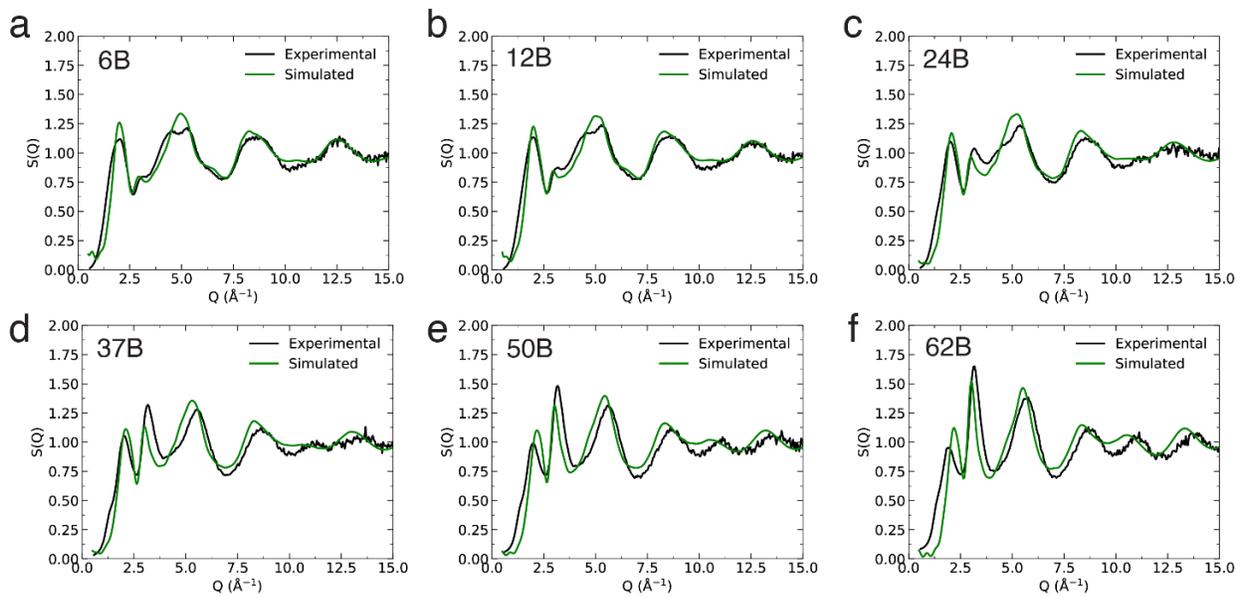

**Figure 6.** Comparison of X-ray structure factors between experimental and simulated borosilicate glasses with (a) 6 mol%, (b) 12 mol%, (c) 24 mol%, (d) 37 mol%, (e) 50 mol%, (f) 62 mol% $B_2O_3$.

## 4. Discussion

### Comparison between multiple parameter sets

To further understand how parameters drive the classic molecular dynamics simulations, and therefore impact the overall simulated structures, we investigate three distinct parameter sets from our optimization process and compare them to our previous work. Namely, (1) Bayesian optimized (this work), which is recommended by Bayesian optimization; (2) CMA-ES optimized, which is recommended by CMA-ES algorithm; (3) $B^4$ optimized, which has the best performance in predicting 4-fold coordinated boron fraction but poor results matching densities; and (4) Wang's potential, which is the starting point for our ML parametrization process. Table 4 shows the B–O and B–B parameters from the Wang's potential (which used incorrect atomic masses)[11], and the corresponding differences (in %) relative to the other parameter sets.

**Table 1.** Comparison of potential parameter sets.

|  | B–O | | | B–B | |
|---|---|---|---|---|---|
|  | $A_{ij}$ (eV) | $\rho_{ij}$ (Å) | $C_{ij}$ (eV·Å$^6$) | $A_{ij}$ (eV) | $\rho_{ij}$ (Å) |
| Wang *et al.* | 206941.81 | 0.1240 | 35.0018 | 484.40 | 0.35 |
| Bayesian optimized | -7.37% | +0.75% | -6.96% | +10% | +0.77% |
| CMA-ES optimized | -3.46% | -0.05% | -23.35% | +3.92% | -6.58% |
| $B^4$ optimized | +14.85% | -1.26% | +2.60% | +4.96% | -14.39% |

## Parameter effect on density and boron speciation

In the following, we assess the accuracy of the various parameter sets by benchmarking simulated glass densities and four-fold coordinated boron fractions against the experimental data. To ensure reliable statistical analysis, we conducted six independent simulations for each composition-parameter combination. Our analysis reveals that both Bayesian and CMA-ES optimization approaches yield parameter sets that accurately reproduce experimental densities and $B^4$ fractions. Notably, despite the algorithmic differences producing distinct optimized parameters as shown in Table 4, the resulting simulation outcomes show no significant variation between these two methods (Fig. 7).

Among all tested parameter sets, the $B^4$-optimized parameters achieved superior accuracy in predicting $B^4$ fractions across the entire compositional range, while other parameter sets exhibited minor systematic deviations. However, all parameter sets successfully captured the characteristic decrease in $B^4$ fraction with increasing $B_2O_3$ content, producing results that align well with experimental trends (Figure 7b). However, a significant limitation emerged for both the $B^4$-optimized parameters and Wang's potential, which substantially underperformed in reproducing experimental density values, especially for boron-rich compositions (Figure 7a). This discrepancy stems from our methodological choice of applying a 1.8 Å cutoff for B–O pair identification during coordination number and $B^4$ fraction calculations. While these parameter sets tend to overestimate glass densities, the simulated B–O bond lengths consistently fall within our specified cutoff range (Figure 8). This explains why $B^4$ fraction calculations remain relatively insensitive to parameter variations, whereas density predictions exhibit much greater sensitivity to parametric changes.

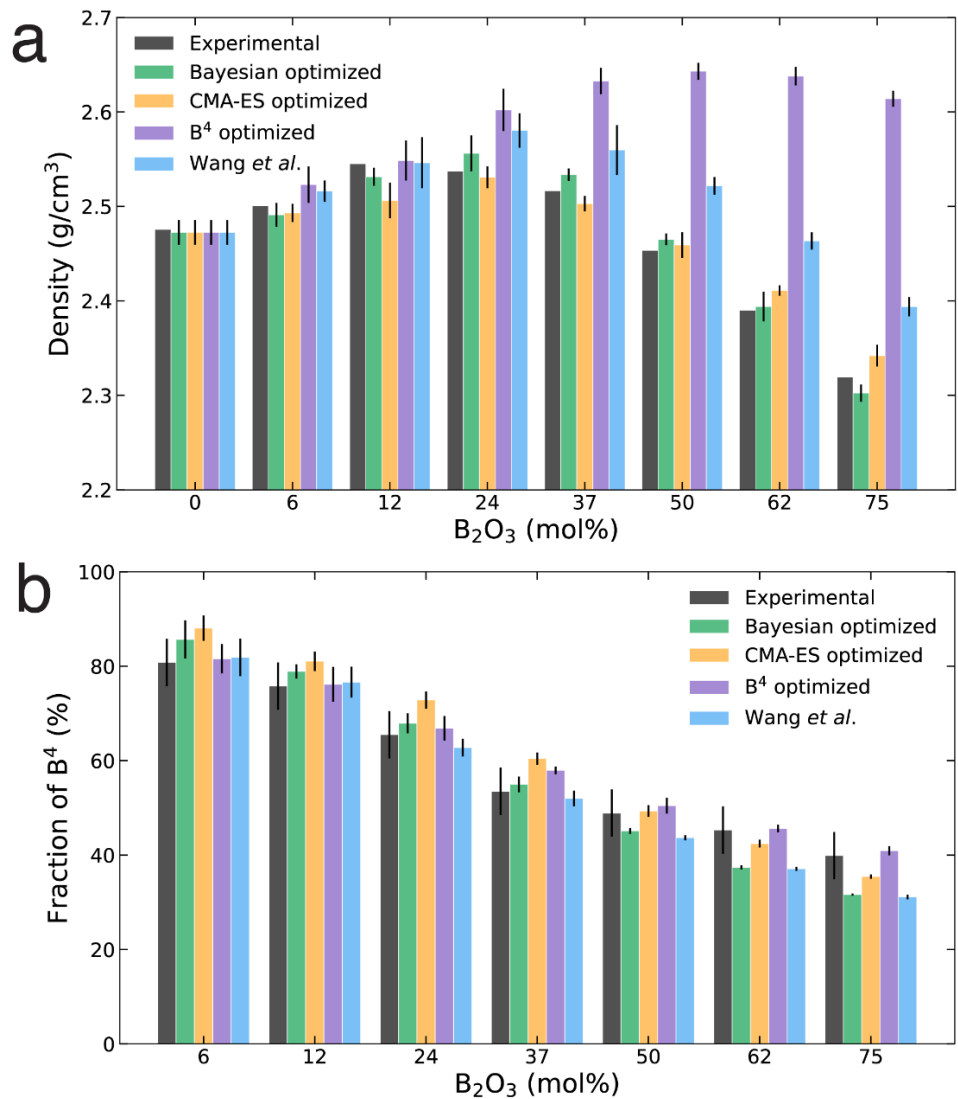

**Figure 7.** Comparison of (a) densities and (b) $B^4$ fractions between experimental data and simulated results for borosilicate glasses with different parameter sets.

## Parameter effect on glass structures

To understand how interatomic parameters influence simulated glass structural properties, we examined the B–O partial PDFs for the different parameter sets (Fig. 8). The positions of O–$B^{III}$ and O–$B^{IV}$ peaks for each parameter set are marked for comparison. All four parameter sets demonstrate a clear relationship between B–O parameter variations and energy calculations, directly affecting bond lengths in both $BO_3$ and $BO_4$ structural units. These bond length changes ultimately govern the structural characteristics of the simulated glasses, providing concrete evidence of the parameter-structure relationship in our computational framework. Notably, despite significant variations in B–O bond length distributions across parameter sets, the calculated $BO_4$ fractions remain remarkably consistent. This stability can be attributed to our 1.8 Å cutoff distance used for B–O pair identification. That is, since the vast majority of B–O bonds fall within this cutoff threshold regardless of the parameter set employed, the coordination number calculations (and consequently the $B^4$ fraction determinations) show minimal sensitivity to these parameter variations.

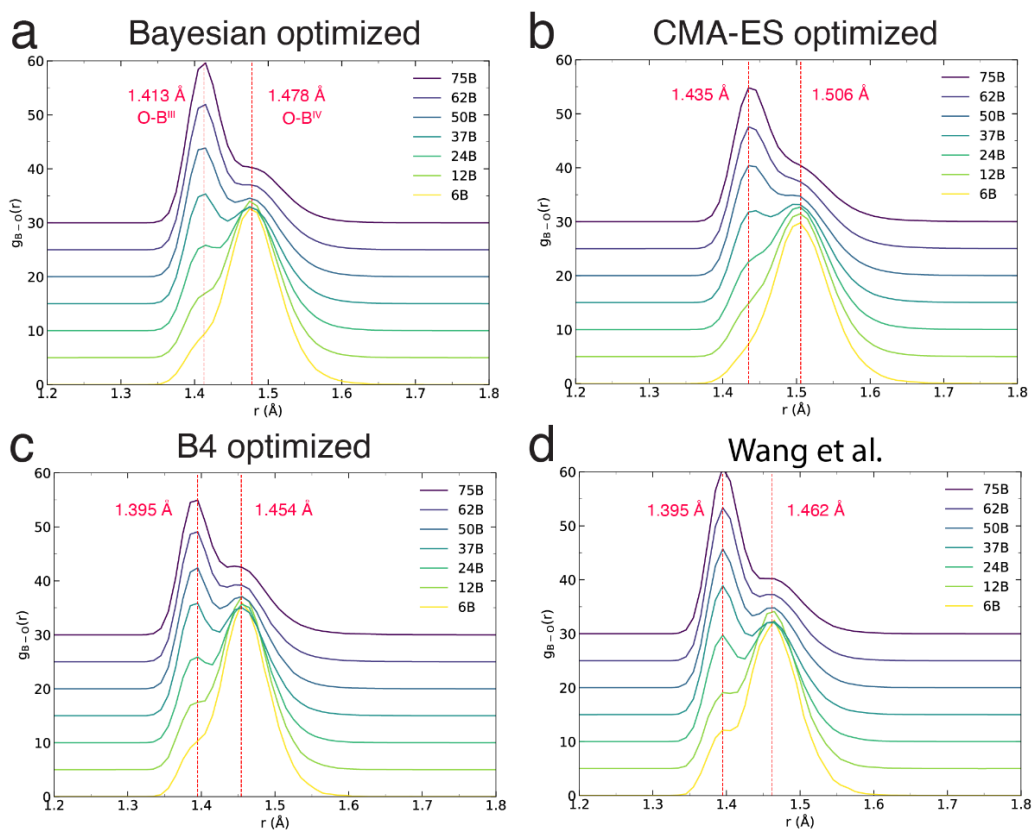

**Figure 8.** B–O partial pair distribution functions of borosilicate glasses simulated by the four different parameter sets: (a) Bayesian optimized, (b) CMA-ES optimized, (c) $B^4$ optimized, and (d) original Wang et al. potential[11]. The dashed lines represent the positions of $B^{III}$–O and $B^{IV}$–O bond peaks.

Figure 9 presents our analysis of the mean X–O–X bond angle distributions (where X = Si or B) obtained from the four different parameter sets, revealing critical insights into their structural accuracy. In detail, the Bayesian optimized, CMA-ES optimized, and Wang's potential parameter sets demonstrate strong performance in reproducing the compositional dependence of mean X–O–X bond angles, successfully capturing the non-monotonic behavior (see dashed trend lines in Figure 9). Conversely, the B$^4$-optimized parameter set exhibits a significant limitation, as does not accurately predict the compositional evolution of X–O–X bond angles. This structural inadequacy provides a direct explanation for the poor density predictions observed across the compositional spectrum. Interestingly, Wang's potential presents a complex case. While it correctly reproduces X–O–X bond angle trends, it nevertheless yields inaccurate density predictions. This is due to the systematically compressed B–O bond lengths in Wang's simulations (Fig. 8), which generate overly compact $BO_3$ and $BO_4$ structural units and result in densities that exceed experimental values.

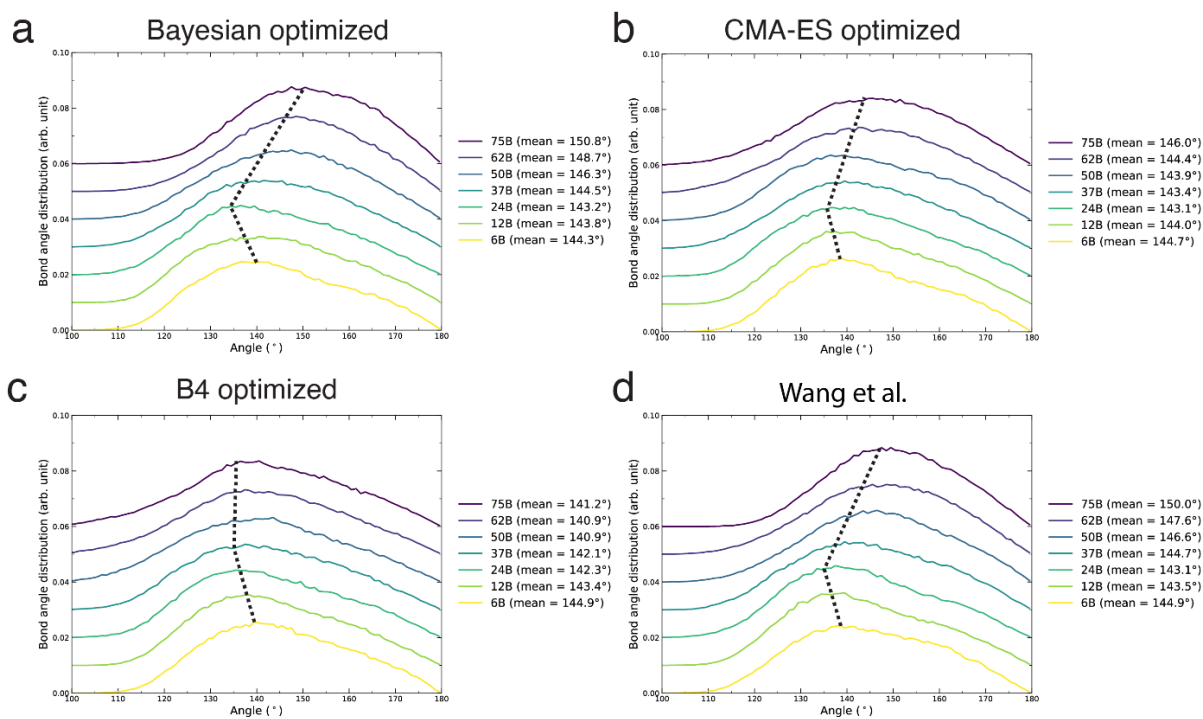

**Figure 9.** Mean X–O–X bond angles (where X = Si or B) of simulated borosilicate glasses for four different parameter sets: (a) Bayesian optimized, (b) CMA-ES optimized, (c) B$^4$ optimized, and (d) original Wang et al. potential[11]. The dashed lines are guides for the eye.

## Relationship between glass density and structure properties

We further investigate how atomic structures affect the simulated glass densities in detail. As discussed in the previous sections, the variations in mean X–O–X bond angle, average B–O bond length, and fraction of BO$_4$ units appear to significantly influence the densities. To this end, Figure 10 shows the simulated densities as a function of mean X–O–X bond angle for different parameter sets with linear regression trendlines. The average B–O bond lengths, which are affected by potential parameters, are labeled near the corresponding trendlines. The marker size represents the fraction of four-fold coordinated boron atoms in the system, with larger marker representing a higher BO$_4$ content.

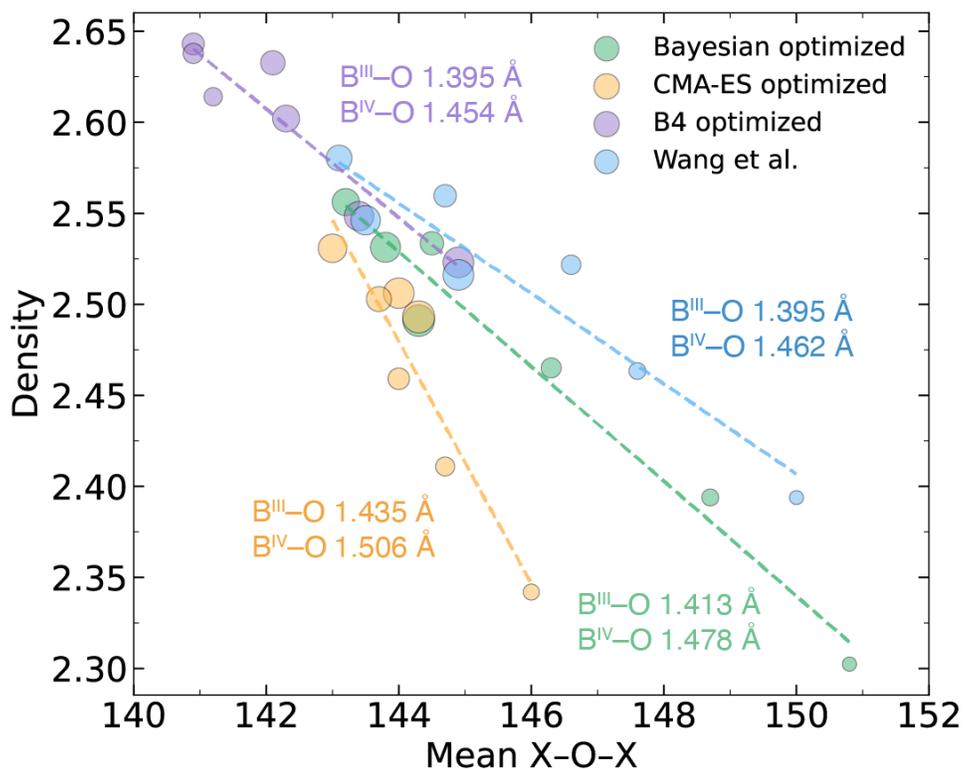

**Figure 10.** Densities as a function of mean X–O–X bond angles for the simulated glasses using four different parameter sets: (a) Bayesian optimized, (b) CMA-ES optimized, (c) B$^4$ optimized, and (d) Wang *et al.* potential[11]. Dashed lines are based on linear regression model. The mean B$^{III}$–O and B$^{IV}$–O bond lengths are reported for each parameter set. The marker size represents the fraction of four-fold coordinated boron atoms, with larger marker representing higher BO$_4$ percentage in the glass.

We observe strong correlation between glass density and mean X–O–X bond angles, suggesting that accurate generation of this structural behavior is essential for reliable density predictions. All parameter sets demonstrate that simulated glass density decreases as mean X–O–X bond angles increase (from left to right in Fig. 10), which is consistent with the physical understanding that larger inter-polyhedral angles correspond to less compact glass structures with lower densities. While B–O (both $B^{III}$–O and $B^{IV}$–O) bond lengths also influence system densities (as longer bonds should result in less dense structures), they represent a secondary effect of lesser importance compared to bond angles. The CMA-ES optimized parameter set produces the longest $B^{III}/B^{IV}$–O bonds, resulting in less compact polyhedral units and consequently generating simulated glasses with the lowest density at any given mean X–O–X bond angle. In contrast, the $B^4$-optimized and Wang's potential parameter sets yield shorter $B^{III}/B^{IV}$–O bond lengths and denser boron polyhedral, producing glasses with higher densities at equivalent mean X–O–X bond angles. This indicates that the $B^{III}/B^{IV}$–O bond length also serves as a key structural parameter controlling the density of simulated glasses. Furthermore, since $BO_4$ units are denser than $BO_3$ units, high $BO_4$ fractions (larger markers) should theoretically result in glasses with high density. This trend is observable for the Bayesian optimized, CMA-ES optimized, and Wang's potential parameter sets, but notably absent in the $B^4$-optimized parameter set. Again, this suggests that the $BO_4$ fraction affects but less strongly correlates with density prediction in simulated borosilicate glasses compared to the more influential factors of $B^{III}/B^{IV}$–O bond length and mean X–O–X bond angles.

## Conclusions

Our work demonstrates the successful development of a machine learning optimized classic potential for molecular dynamics simulations, which addresses the longstanding challenges in accurately simulating borosilicate glasses across diverse compositions. The potential's transferability and ability to reproduce experimental structural data, while being optimized on only two key properties, highlights the power of targeted machine learning approaches in materials science. By systematically investigating how empirical force field parameters influence macroscopic glass properties, this study provides crucial insights into the fundamental connections between atomic-scale interactions and bulk material behavior. These findings not only advance our computational capabilities for modeling complex glass systems but also establish a framework for understanding and predicting structure-property relationships in borosilicate glasses. The methodology presented here opens new avenues for accelerated design and optimization of glass compositions with tailored properties for specific applications.

## Acknowledgement

M.M.S. acknowledges funding from the European Union (ERC, NewGLASS, 101044664) and Carlsberg Foundation (CF23-0958).

## CRediT authorship contribution statement

**Kai Yang**: Writing – original draft, Methodology, Investigation, Visualization, Data curation, Formal analysis. **Ruoxia Chen**: Methodology, Investigation, Data curation. **Anders K.R. Christensen**: Writing – review & editing, Investigation, Data curation. **Mathieu Bauchy**: Conceptualization, Methodology. **N.M. Anoop Krishnan**: Writing – review & editing, Supervision. **Morten M. Smedskjaer**: Writing – review & editing, Methodology, Resources, Supervision. **Fabian Rosner**: Writing – review & editing, Investigation, Supervision.

## Data Availability

The data supporting the findings in this paper are available at https://github.com/KaiYang1010/MD-Borosilicate-potential.

## Declaration of competing interest

Given his role as Editor, Morten M. Smedskjaer had no involvement in the peer review of this article and had no access to information regarding its peer review. Full responsibility for the editorial process for this article was delegated to another journal editor. If there are other authors, they declare that they have no known competing financial interests or personal relationships that could have appeared to influence the work reported in this paper.

## Declaration of generative AI and AI-assisted technologies in the writing process

During the preparation of this work, the authors used Claude 3.7 Sonnet in order to improve the readability and language of the manuscript. The authors reviewed and edited the content as needed and take full responsibility for the content of the published article.